\documentclass[preprint]{revtex4}
\usepackage[T1]{fontenc}
\usepackage{graphicx}
\usepackage{rotating}
\usepackage{bm}        
\usepackage{amssymb}   
\usepackage{bm}
\def\etal{{\it et al.}}

\def\jpc#1#2#3{J.~Phys.~Chem.~{\bf #1},\ #2\ (#3)}
\def\jcp#1#2#3{J.~Chem.~Phys.~{\bf #1},\ #2\ (#3)}
\def\cpl#1#2#3{Chem.~Phys.~Lett.~{\bf #1},\ #2\ (#3)}

\def\pra#1#2#3{Phys.~Rev.~A~{\bf #1},\ #2\ (#3)}
\def\prl#1#2#3{Phys.~Rev.~Lett.~{\bf #1},\ #2\ (#3)}

\def\jpb#1#2#3{J. Phys. B: At. Mol. Opt. Phys. {\bf #1},\ #2\ (#3)}
\def\rmp#1#2#3{Rev.~Mod.~Phys.~{\bf #1},\ #2\ (#3)}

\def\ddpar{\partial}

\def\LA{L_{\rm A}}
\def\LB{L_{\rm B}}

\def\k1{k_1}
\def\k2{k_2}
\def\q1{q_1}
\def\q2{q_2}
\newcommand{\beq}{\begin{equation}}
\newcommand{\eeq}{\end{equation}}

\begin{document}
\date{\today}
\flushbottom \draft
\title{
Molecules Near Absolute Zero and External Field Control of Atomic and Molecular Dynamics
}
\author{R. V. Krems}
\affiliation{
Department of Physics, Harvard University,
Cambridge, MA 02138
\footnote{The author is part of the Harvard-MIT Center for Ultracold Atoms and the Institute for Theoretical Atomic, Molecular and Optical Physics at the Harvard-Smithsonian Center for Astrophysics; Electronic mail: rkrems@cfa.harvard.edu}}

\centerline{\bf \small published in Int. Rev. Phys. Chem. vol 24, pp.99-118 (2005)}

\begin{abstract}
This article reviews the current state of the art in the field of cold and ultracold 
molecules and demonstrates that chemical reactions, inelastic collisions and dissociation 
of molecules at subKelvin temperatures can be manipulated with 
external electric or magnetic fields. 
The creation of ultracold molecules may allow for spectroscopy measurements with extremely high precision and tests of fundamental symmetries of nature, quantum computation with molecules as qubits, and controlled chemistry. The probability of chemical reactions and collisional energy transfer can be very large at temperatures near zero Kelvin. The collision energy of ultracold atoms and molecules is much smaller than perturbations due to interactions with external electric or magnetic fields 
available in the laboratory. External fields may therefore be used to induce dissociation of weakly bound molecules, stimulate forbidden electronic transitions, suppress the effect of centrifugal barriers in outgoing reaction channels or tune Feshbach resonances that enhance chemical reactivity. 

\end{abstract}

\maketitle

\clearpage
\newpage

\section{Introduction}

Controlling chemical reactions with electromagnetic 
fields has long been a sought-after goal 
of researchers. 
External field control of chemical reactions will not only allow chemists to selectively 
produce desired species, but also reveal mechanisms of chemical reactions, 
yield information on interactions determining chemical reactions and elucidate 
the role of non-adiabatic and relativistic effects in chemical dynamics. 
Possible applications
of controlled chemistry range from quantum 
computation with molecules, to fundamental 
tests of reaction rate theories, to studies 
of fine details of molecular structure or intermolecular interaction potentials. 
External fields may influence molecular collisions 
when the translational energy of 
the molecules is smaller than the perturbation due to 
interactions with external fields. 
Moderate magnetic and electric fields available in the laboratory 
shift molecular energy levels by up to a few Kelvin 
so external field control of molecular dynamics in the gas phase is most easily achieved  
at temperatures near or less than one Kelvin \cite{note1}.

A variety of methods have been developed in the past seven years to 
create molecules at temperatures  between 
$10^{-3}$ and $1$ K  (cold molecules)
and at temperatures between $10^{-9}$ and $10^{-3}$ K 
(ultracold molecules). The experimental techniques for the production of cold 
and ultracold molecules have been described in a recent 
review by Doyle and coworkers \cite{editorial}. Table 1 lists the coldest molecules produced 
in laboratory studies to date. 

 Ultracold molecules are characterized by unusual scattering properties. The orbital angular momentum of the colliding molecules is asymptotically zero \cite{note2}
  and the collision dynamics is determined by Wigner's threshold laws 
 \cite{wigner} at such low temperatures. Cross sections for inelastic scattering and chemical reactions 
 rise to infinity as the collision energy vanishes so that reaction rate constants are independent of 
 temperature and they can be quite large in the limit of zero Kelvin. 
Collisions of ultracold molecules are extremely sensitive to intermolecular interaction potentials
and relative energies of the initial and final scattering states. A slight variation of the molecular structure 
due to an applied external field may, therefore,  dramatically change the outcome of an inelastic collision or chemical reaction of ultracold molecules.

The purpose of this article is to review the current state of the art in 
the field of ultracold molecules and demonstrate that dynamics of ultracold 
molecules can be manipulated with 
external electric or magnetic fields. 
I will show that external field control of molecular dynamics can be based on several principles. 
Zeeman and Stark effects may remove some of the energetically allowed reaction 
paths or they may open closed reaction channels, leading to suppression or 
enhancement of the reaction efficiency \cite{krems2004}. External fields couple 
the states of different total angular momenta, so that forbidden electronic 
transitions may become allowed in an external field and the transition rate 
may be controlled by the field strength \cite{krems2003}.  
The rate of low temperature abstraction reactions may be dramatically enhanced by 
the presence of a resonance state near threshold \cite{bodo2004}. 
By shifting molecular energy levels with external fields, it should be possible 
to bring an excited bound level of the reactive complex in resonance 
with the collision energy. 
Finally, external fields may influence 
statistical properties of ultracold molecular gases such as diffusion.

This introduction is followed by a brief discussion of Wigner's threshold laws and 
 methods for the production of cold and ultracold molecules. A particular accent is 
 made on the possibility of the creation of ultracold molecules in other than 
 $\Sigma$-symmetry electronic  states and magnetic trapping of atoms in states of non-zero electronic orbital angular momenta. 
 Quantum mechanical theory of molecular dynamics in external fields is outlined in 
 the next section and mechanisms for external field control of molecular dynamics are 
 described in the final section before conclusions. 
 In preparing this paper, I made an effort to avoid overlap with previous recent reviews 
\cite{krems2002review,bethlem2003,editorial}.
In particular, the editorial review of Doyle 
\etal~\cite{editorial} described the experimental work with cold and ultracold molecules so 
the present discussion is primarily about recent theoretical work.

\section{Reactions at zero temperature}
 
 Wigner \cite{wigner}  showed that elastic scattering ($\sigma_{\rm el}$) and 
 reaction ($\sigma_{\rm r}$) cross sections vary in the limit of 
 vanishing collision velocity ($v$) as
  \begin{eqnarray}
 \sigma_{\rm el}(l)  \sim v^{4l}
 \label{elastic_crosssection}
 \\
 \sigma_{\rm r}(l) \sim v^{2l-1}
\label{reaction_crosssection} 
\end{eqnarray}
 
 \noindent
 where $l$ is the angular momentum for orbital motion of the colliding molecules about each 
 other. Eqs. (\ref{elastic_crosssection}) and (\ref{reaction_crosssection})
are valid for systems with short-range interactions. Long-range intermolecular interactions may modify the 
threshold laws (see, e.g., the review of Sadeghpour \etal~\cite{hossein}). 
It follows from the Wigner expressions that both the elastic scattering and chemical reactions are dominated by collisions with $l=0$ at ultracold temperatures. 
The elastic cross section is thus independent of the collision energy and the reaction cross section is inversely proportional 
to the velocity, rising to infinity as $v \rightarrow 0$.
Eq. (\ref{reaction_crosssection}) with $l=0$ is valid for inelastic collisions or chemical reactions 
 providing the intermolecular interaction potential decreases with intermolecular separation $R$ faster than $1/R^2$ in the long range. 
A simple derivation 
of the expressions (\ref{elastic_crosssection}) and (\ref{reaction_crosssection}) 
was given by Landau and Lifshitz \cite{landau}.

  The reaction rate constant is obtained from the energy dependence of the reaction cross section as follows
    \begin{eqnarray}
  K = \left ( \frac{8 k_b T}{\pi \mu} \right )^{1/2} \int_{0}^{\infty} \sigma_{\rm r}(E) e^{-E/k_BT} \frac{E {\rm d}E}{(k_bT)^2},
\label{rate}
\end{eqnarray}

\noindent
where $k_B$ is the Boltzmann constant, $\mu$ is the reduced mass of the reacting 
molecules, $T$ is the temperature and $E = \mu v^2/2$. Replacing $\sigma_{\rm r}$ by $1/\sqrt{E}$ in Eq. (\ref{rate})
yields
\begin{eqnarray}
K  \sim {\rm const}.
\end{eqnarray}

\noindent
The reaction rate constant is thus independent of temperature and finite in the limit $T \rightarrow 0$~K. 
Following the analysis of Landau and Lifshitz \cite{landau} and Mott and Massey \cite{mott},
Balakrishnan \etal~\cite{bala1997} and Bohn and Julienne \cite{bohn1997} 
showed that it is most convenient to express the zero temperature 
reaction rate in terms of the imaginary part of the scattering length related to off-diagonal elements of the scattering $S$-matrix.

Several recent calculations of ro-vibrational relaxation  in atom - molecule collisions 
\cite{bala1998,forrey1999,bala2000,bala2001,zhu2001,flasher2002,bodo2002,bala2003,florian2004,tilford2004},
 non-adiabatic electronic relaxation in atom - atom and atom - molecule collisions 
 \cite{krems2001,siska2001,krems2002,krems2002a}
  and reactive scattering 
\cite{bala2001a,bala2003a,bodo2002a,bodo2002b,soldan2002,cvitas2005,cvitas2005a,quemener2005}
  showed that 
rate constants for inelastic energy transfer and chemical reactions have significant magnitudes at zero Kelvin.  In particular, Balakrishnan and Dalgarno \cite{bala2001a} found that  the chemical reaction F $+$  H$_2$  $\rightarrow$ HF $+$ H occurs very rapidly at ultracold temperatures despite a large activation barrier of about $1.5$ kcal/mol.  The rate constant for this reaction was calculated to be as large as $1.25 \times 10^{-12}$  cm$^3$ sec$^{-1}$ at zero Kelvin.
Sold\'{a}n \etal~\cite{soldan2002} 
and Cvita\v{s} \etal~\cite{cvitas2005a}
showed that chemical reactions without activation barriers (insertion reactions)
are even more efficient at ultralow energies. The results of their extensive  calculations based 
on a hyperspherical coordinate representation of the wave function yielded the zero temperature 
rate constant $5 \times 10^{-10}$ cm$^3$ sec$^{-1}$
for the Na + Na$_2(v>0)$ reaction and $4.1 \times10^{-12}$ cm$^3$ sec$^{-1}$ for 
the $^7$Li + $^6$Li$^7$Li$(v=0)$ reaction.

\section{Methods to create ultracold molecules}

An experiment on cooling atoms to ultracold temperatures includes two stages: trapping and evaporative cooling of trapped atoms \cite{ketterle}. Atoms are  precooled by laser cooling and captured in a magneto-optical or purely magnetic trap \cite{ketterle,phillips}. 
Trapping isolates the atomic cloud from the thermal environment. 
The magnetic trap is a superposition of magnetic fields that creates a minimum of the magnetic field potential at the center of the experimental cell. Magnetic traps confine atoms in Zeeman states with a positive gradient of energy with respect to the field strength (low-field seeking states). The potential energy of atoms in low-field seeking states varies in a magnetic trap as shown by the full line in Fig. 1. 
It is essential for a trapping experiment that atoms remain in the low-field seeking state upon collisions. 
If the orientation of atomic magnetic moments with respect to the magnetic field axis is changed, the trapped atoms relax to a high-field seeking state whose potential energy varies in the trap as shown by the broken line in Fig.~1. Atoms in high-field seeking states are not trappable. The density of trapped atoms is largest near the center of the trap, where the translational temperature of the atomic cloud is lowest. 
Collisional relaxation to high-field seeking states, therefore,  removes the coldest atoms and leads to heating. 

The evaporative cooling  is a repetitive process of driving most energetic atoms out of the trap and re-equilibrating the kinetic energy of the remaining atoms. 
The evaporative cooling is an integral part of any experiment with ultracold atoms and it rests on 
the efficiency of energy transport in elastic collisions.  

Laser cooling is applicable to a limited number of specific atoms. An alternative and more general method of trapping atoms relies on buffer-gas loading \cite{doyle1995}. Hot atoms are introduced into a cell of cold buffer gas, usually $^3$He, and cooled by elastic collisions with the buffer gas atoms. When the temperature of the atoms becomes less than 1 K, they can be trapped in a large-gradient magnetic trap. The buffer gas can then be pumped out and the trapped atoms can be cooled to ultralow temperatures by evaporative cooling \cite{weinstein2002}.

Laser cooling cannot be applied to molecules (see, however, the work of Di Rosa \cite{dirosa2004}). A variety of alternative methods have been developed to precool and trap molecules. An incomplete list includes buffer-gas loading \cite{weinstein1998}, Stark deceleration \cite{bethlem2003}, skimming
\cite{rangwala2003,nikitin2003},  mechanical slowing \cite{gupta1999}
and crossed-beam collision \cite{elioff2003}. Trapped molecules can be evaporatively cooled to ultralow temperatures. In addition, ultracold molecules can be created by linking ultracold atoms together as in photoassociation or through Feshbach resonances (see \cite{editorial} and references therein). The photoassociation and Feshbach resonance methods 
produce ultracold molecules directly so they do not depend on the evaporative cooling of molecules which may be difficult to implement. The review of Doyle and coworkers \cite{editorial} presents a more comprehensive discussion and references of methods for the production of ultracold molecules. 

  The stability of molecules in a magnetic trap was studied by several authors 
\cite{bohn2000,bohn2000a,bohn2001,avdeenkov2001,volpi2002,volpi2002a,krems2003a,krems2003b,krems2004a,groenenboom2005}.
   Bohn and coworkers demonstrated that molecules without electronic orbital angular momenta ($\Sigma$-state molecules) tend to preserve the orientation of the magnetic moment in a magnetic field
   \cite{bohn2000,bohn2000a,bohn2001,avdeenkov2001,volpi2002,volpi2002a}. Krems and coworkers showed that the Zeeman relaxation in rotationally ground-state molecules without electronic orbital angular momenta is mediated by coupling to rotationally excited molecular states \cite{krems2003a,krems2003b,krems2004a}. Collisional trap loss is therefore sensitive to the rotational constant of trapped molecules and Krems and Dalgarno concluded that only $\Sigma$-state molecules with {\it large} rotational constants are amenable to evaporative cooling in a magnetic trap \cite{krems2004a}. 
A recent experiment of  Maussang \etal~confirmed these predictions \cite{maussang2005}. 

The interaction potential anisotropy in molecule - molecule collisions is larger than that in collisions of molecules with He atoms. It should be expected that collisional spin depolarization will be more efficient in molecule - molecule collisions relevant for evaporative cooling experiments and the possibility of buffer-gas loading of molecules in a magnetic trap does not guarantee the possibility of evaporative cooling of molecules. However, Krems and Dalgarno pointed out that the mechanism of trap loss in molecule - molecule collisions is the same as in atom - molecule collisions, providing weak magnetic dipole interactions  can be neglected \cite{krems2004a}. The basic result that $\Sigma$-state molecules with larger rotational constants will be more stable in magnetic traps, therefore, applies to evaporative cooling experiments as well. The quantitative question of whether the evaporative cooling of molecules is going to be possible in a magnetic trap remains open.

Molecules with dipole moments can potentially be trapped in electrostatic traps.  However, Bohn found that collisional relaxation of molecules from low-electric-field-seeking to high-electric-field-seeking states precludes the electrostatic trapping \cite{bohn2001}. Bohn recommended that magnetic fields should be used for trapping polar paramagnetic molecules. 

  Proposals exist  \cite{meijer} to cool molecules to extremely low temperatures by collisions with ultracold trapped alkali metal atoms (sympathetic cooling). Sold\'{a}n and Hutson explored the possibility of cooling NH molecules by collisions with ultracold Rb atoms \cite{soldan2004}. They found that cooling may be complicated by non-adiabatic dynamics  involving ion-pair states and leading to dramatic enhancement 
of the interaction strength. The authors proposed an experiment to create strongly bound triatomic complexes at ultracold temperatures.

Molecular Bose-Einstein condensates have recently been produced by coupling ultracold atoms through magnetically tuned Feshbach resonances \cite{greiner2003,jochim2003,zwierlein2003}.
  Particularly successful were the experiments with fermionic ultracold atoms. 
Petrov \cite{petrov2003} and Petrov \etal~\cite{petrov2004} showed that weakly bound molecules composed of fermionic atoms are stable against collisionally induced relaxation and trap loss due to Fermi suppression in atom - atom interactions. This raised the question of whether vibrational relaxation in strongly bound dimers of fermionic
atoms would
also be suppressed in ultracold collisions. 
Cvita\v{s} \etal~computed the vibrational relaxation efficiency in collisions of Li with Li$_2$ for both bosonic and fermionic atoms \cite{cvitas2005}. The authors found that the rate constants for vibrational relaxation of Li$_2$ in the lowest three vibrationally excited levels are
similar for bosonic and fermionic atoms so the quenching rates are not suppressed in strongly bound molecules composed 
of fermionic atoms. 

Ultracold  collisions are very sensitive to intermolecular interaction potentials; in particular, Sold\'{a}n \etal~\cite{soldan2003} demonstrated that three-body non-additive 
forces between spin-polarized alkali atoms are crucial for dynamics of alkali atom - alkali molecule reactions. 
An important question yet to be answered is whether the accuracy of modern quantum chemistry methods
is sufficient for a realistic {\it ab initio}  description of molecular dynamics at ultracold temperatures. 

The emphasis of several recent experiments was on the creation of ultracold polar molecules. 
The work of Bergeman \etal~\cite{bergeman2004} on the creation of translationally, electronically, vibrationally and rotationally cold 
RbCs molecules and the work of Wang \etal~\cite{wang2004} on the photoassociative molecule formation and trapping of KRb 
molecules are two most recent examples of the progress in this direction.

  Dynamics of molecules with non-zero electronic orbital angular momenta about the molecular axis (non-$\Sigma$-state molecules) is characterized by non-adiabatic interactions 
 and it can be manipulated with external electric and magnetic fields \cite{ticknor2005}. The creation of ultracold non-$\Sigma$-state molecules will open up possibilities for studies of non-adiabatic effects in chemical reactions at zero Kelvin. High precision spectroscopy  measurements of non-$\Sigma$-state  molecules possible at ultracold temperatures may allow for tests of fundamental symmetries of the nature and help in searches for the time variation of fundamental constants. Trapping and cooling non-$\Sigma$-state molecules to ultracold temperatures thus presents a particular interest.  Avdeenkov and Bohn \cite{avdeenkov2002} and Groenenboom \cite{groenenboom2005} found that evaporative cooling of non-$\Sigma$-state molecules in a magnetic trap will unlikely be possible due to large collisional losses. Alternatively, ultracold non-$\Sigma$-state molecules can be created by photoassociation of atoms with non-zero electronic orbital angular momenta. However, all atoms cooled to ultracold temperatures in a magnetic trap so far were in electronic $S$-states and the electronic ground state of molecules produced by photoassociation of  $S$-state atoms is always of $\Sigma$ symmetry. An important question is thus ``Can atoms with non-zero electronic orbital angular momenta (non-$S$-state atoms) be magnetically trapped and evaporatively cooled to ultracold temperatures?''

 \section{Magnetic trapping of non-$S$-state atoms}

 An interaction of two atoms with non-zero electronic orbital angular momenta gives rise to several 
 adiabatic potentials with different symmetries. For example, the interaction of two oxygen atoms in the ground $^3P$ state is described by 18 interaction potentials corresponding to  
 different spin multiplicity, 
 parity and spatial symmetry of the O$_2$ molecular states \cite{zygelman1994}. 
 A simpler complex  of O($^3P$) with He($^1S$) is described by two interaction potentials of $\Sigma$ and $\Pi$ symmetries \cite{krems2002b}.

 Krems, Groenenboom and Dalgarno \cite{krems2004b} showed that the electronic interaction between two atoms in arbitrary angular momentum states can be expressed in an effective potential form as 
 
 \begin{eqnarray}
\nonumber
\hat{V}^S =  (4 \pi)^{1/2} \sum_{k_1} \sum_{k_2} \sum_{k} V^S_{k_1,k_2,k}(R) 
\sum_{q_1} \sum_{q_2} \sum_{q} (-1)^{k_1 - k_2} 
  \left ( \begin{array}{c c c} 
                                                  k_1   & k_2     &  k  \\
                                                  q_1   & q_2     &  q  \\
                              \end{array} \right )
\hat{T}^{k_1}_{q_1}(\LA) \hat{T}^{k_2}_{q_2}(\LB) Y_{k q}(\hat{R}),
\\
\label{TheExpansion}
\end{eqnarray}

\noindent
where big parentheses denote $3j$-symbols, $\hat{T}^{k}_q$ are spherical 
tensors describing the unpaired electrons in the separated 
open-shell atoms, $\LA$ and $\LB$ are the electronic orbital angular momenta of the two atoms (denoted A and B), $S$ is the total electronic spin of the system, and  $V^S_{k_1,k_2,k}(R)$
are some coefficients related to the adiabatic interaction potentials of the diatomic molecule.

Eq. (\ref{TheExpansion}) can be used to define the electronic interaction anisotropy between atoms in arbitrary electronic states. The terms with $k_1=k_2=k=0$ represent the isotropic part of the electronic interaction that cannot induce angular momentum transfer in atom - atom collisions. The terms with non-zero $k_1$ or $k_2$ and $k$ represent the anisotropic part of the electronic interaction. As follows from the results of Ref. \cite{krems2004b}, the anisotropic coefficients $V^S_{k_1, k_2, k}$ can all be expressed in terms of differences between the adiabatic potentials of the diatomic molecule. For example, the electronic interaction anisotropy in the O($^3P$) -- He
 complex is
 determined by the splitting of the $\Sigma$ and $\Pi$ interaction potentials at finite interatomic distances  - the result obtained much earlier by Aquilanti and Grossi \cite{aquilanti}.

The stability of atoms in a magnetic trap is determined by the degree of the electronic interaction anisotropy in atomic collisions. The non-relativistic interaction of $S$-state atoms is isotropic. All terms with non-zero $k_1$, $k_2$ or $k$ in Eq. (\ref{TheExpansion}) vanish when both interacting atoms are in states with zero electronic orbital angular momentum.  Collisional relaxation and trap loss of $S$-state atoms in maximally stretched low-field seeking states is determined by the magnetic dipole-dipole interaction which is very weak. Collisional relaxation and trap loss of non-$S$-state atoms may, by contrast, be induced by the anisotropy of the electrostatic interaction, {\it i.e.} by terms with non-zero $k_1$, $k_2$ and $k$ in Eq. (\ref{TheExpansion}).

To understand the prospects for buffer-gas loading and evaporative cooling of non-$S$-state atoms, Krems and Dalgarno  carried out rigorous quantum mechanical calculations of rate constants for elastic scattering and Zeeman relaxation in collisions of  oxygen atoms in the $^3P$ state with 
$^3$He atoms \cite{krems2003}. It was found that the probability of elastic He - O($^3P$) collisions is only six times larger than the probability of Zeeman relaxation at the temperature $1$ K and magnetic field $1$ Tesla. The buffer-gas loading and evaporative cooling experiments are possible when the  elastic scattering rate exceeds the rate for trap loss by, at least, one thousand.  
The results of Krems and Dalgarno thus indicated that buffer-gas loading and evaporative cooling of non-$S$-state atoms like oxygen would be impossible.  
Kokoouline \etal~\cite{kokoouline2003} arrived at the same conclusion in their numerical study of Sr($^3P$) - Sr($^3P$) collisions.

Hancox and coworkers \cite{cindy} and Krems and coworkers \cite{krems2005} have recently reported a combined experimental and theoretical study of the electronic interaction anisotropy in collisions of transition metal atoms Sc and Ti with $^3$He. It was found that the interaction anisotropy in complexes of Sc($^2D$) and Ti($^3F$)  with He is dramatically suppressed due to the presence of paired spherically-symmetric electrons in the outer electronic shell of the transition metal atoms. 
Non-relativistic interaction potentials for complexes of Ti and Sc atoms with He were described in detail
in Ref. \cite{jacek}.
Table II presents the ratio of rate constants for elastic scattering and Zeeman relaxation in collisions 
of oxygen in the $^3P$ and $^1D$ states with $^3$He and in collisions of the transition metal atoms with  $^3$He. While O($^3P$) and O($^1D$) atoms tend to change the orientation of their magnetic moment in almost every other collision, 
only one in about 40000 collisions of Ti($^3F$) with $^3$He leads to angular momentum transfer and loss of magnetically trapped Ti. 
Hancox \etal~\cite{cindy_nature} have later found that the interaction anisotropy is even more suppressed in complexes of rare-earth atoms with He and most of the rare-earth atoms have been magnetically trapped as a result of this work. 

Interactions between transition metal atoms or rare-earth atoms are more complex than those between transition metal atoms and He. The work outlined above rules out the possibility of evaporative cooling of main-group non-$S$-state atoms like oxygen or strontium in a magnetic trap, but provides only an indication  that the evaporative cooling of non-$S$-state transition metal and rare-earth atoms may be possible. It should be expected, however, that interactions between transition metal atoms and $S$-state alkali metal atoms in a magnetic trap will be similar to those between transition metal atoms and He and the results of Hancox \etal~ and Krems \etal~suggest the possibility of sympathetic cooling of the non-$S$-state atoms by collisions with trapped alkali metal atoms to ultracold temperatures. 
Ultracold non-$S$-state atoms may be photoassociated with each other or with the alkali metal atoms to produce ultracold non-$\Sigma$-state molecules.

\section{Collisions of molecules in external fields}

  Following the work of Volpi and Bohn \cite{volpi2002}, Krems and Dalgarno \cite{krems2004a,krems2004c} presented a quantum mechanical theory of molecular collisions in external fields. External electric and magnetic fields disturb the spherical symmetry of the problem and the total angular momentum of the colliding molecules is not a good quantum number in the presence of 
  an external field. The total angular momentum representation of Arthurs and Dalgarno \cite{alex} does not reduce the dimension of the collision problem in an external field and Krems and Dalgarno \cite{krems2004a,krems2004c} argued that the most convenient theory of molecular collisions should then be based on a {\it fully uncoupled space-fixed} basis representation of the wave function. 
The complexity of the collision theory in the fully uncoupled representation does not increase with the number of internal degrees of freedom and the evaluation of the matrix elements of all terms in the Hamiltonian  is straightforward. 
  
  The Hamiltonian of two $\Sigma$-state molecules (A and B) with non-zero spins in an external magnetic field can be generally   written as

\begin{eqnarray}
H = - \frac{1}{2 \mu R} \frac{\ddpar ^2}{\ddpar R^2} R + \frac{{\bf l}^2}{2 \mu R^2} + U
\label{ham}
\end{eqnarray}

\noindent
where $R$ is the Jacobi coordinate joining the 
centers of mass of the colliding particles,
${\bf l}$ is the angular momentum describing  
the rotation of the vector ${\bf R}$, $\mu$ is the reduced
mass of the colliding molecules,

\begin{eqnarray}
U =  H^{\rm si}_{\rm A} + H^{\rm si}_{\rm B}
+ V_{\rm A}^{\rm sd} + V_{\rm B}^{\rm sd}
+   V_{\rm AB}^{\rm si} + V_{\rm AB}^{\rm sd},
\label{matrixU}
\end{eqnarray}

\noindent
$H_{\rm A}$ and $H_{\rm B}$ denote the Hamiltonians of the isolated 
molecules A and B at zero field, $V_{\rm AB}$ denotes the
potentials for the interaction between A and B 
and the superscripts ``${\rm si}$''
and ``${\rm sd}$'' distinguish the spin-independent (${\rm si}$)
and spin-dependent (${\rm sd}$) terms. 
The terms $ V_{\rm A}^{\rm sd}$ and $V_{\rm B}^{\rm sd}$
describe the interaction with external magnetic fields. 
  
The total wave function of the colliding molecules is 
expanded  
in products of eigenfunctions of ${\bf l}^2$, $H^{\rm si}_{\rm A}$, 
$H^{\rm si}_{\rm B}$, ${\bf S}^2_{\rm A}$ and ${\bf S}^2_{\rm B}$ \cite{krems2004a}:

\begin{eqnarray}
\Psi = {R}^{-1} \sum_{i} F_{i}(R) \phi_{i},
\label{psi}
\end{eqnarray}
\noindent
where 
\begin{eqnarray}
\phi_{i} = \phi_{\rm A} \phi_{\rm B} |S_{\rm A} M_{S_{\rm A}} {\rangle} 
|S_{\rm B} M_{S_{\rm B}} {\rangle} |l m_l {\rangle}, 
\label{expansion}
\end{eqnarray}

\noindent
$S_{A}$ and $S_{B}$ are the spins of the molecules ${\rm A}$ and 
${\rm B}$, and $m_l$, $M_{S_{\rm A}}$ and $M_{S_{\rm B}}$ are 
the projections of ${\bf l}$, ${\bf S}_{\rm A}$ and ${\bf S}_{\rm B}$
on the magnetic field axis. The functions $\phi_{\rm A}$ and 
$\phi_{\rm B}$ are obtained from the eigenstate equations with 
the Hamiltonians $H^{\rm si}_{\rm A}$ and $H^{\rm si}_{\rm B}$. 

The matrix of the Hamiltonian (\ref{ham}) is not diagonal in the basis (\ref{expansion}) in the limit
$R \rightarrow \infty$ and the scattering $S$-matrix cannot be found in this representation. 
An additional transformation ${\bf C}$ must be introduced to form a basis in which the Hamiltonian of the separated molecules would be diagonal. The matrix elements of this transformation cannot be found analytically, even though the structure of the matrix ${\bf C}$ is determined by the 
matrices ${\bf H}^{\rm si}_{\rm A}$, ${\bf V}_{\rm A}^{\rm sd}$, ${\bf H}^{\rm si}_{\rm B}$
and ${\bf V}_{\rm B}^{\rm sd}$. The matrix of the the  transformation 
${\bf C}$ is constructed from eigenvectors of the asymptotic Hamiltonian

\begin{eqnarray}
H_{\rm as}  = H^{\rm si}_{\rm A} + H^{\rm si}_{\rm B}
+ V_{\rm A}^{\rm sd} + V_{\rm B}^{\rm sd}
\label{asymptotic}
\end{eqnarray}

\noindent
and it should be recomputed at every value of the external field. 
The eigenstates of the ${\bf C^{\rm T} H_{\rm as} C}$
matrix labeled by the indexes $\alpha, l, m_l$ are the scattering channels of 
the colliding molecules in the presence of the external field. 

The interaction of the orbital angular momentum ${\bf l}$ with magnetic fields is neglected so both
the matrix of the asymptotic Hamiltonian (\ref{asymptotic}) and the ${\bf C}$ matrix are diagonal in $l$ and $m_l$ quantum numbers. This approximation allows us to apply the following boundary conditions 

\begin{eqnarray}
F^{\alpha lm_l}_{\alpha'l'm_l'}(R \rightarrow 0) \rightarrow 0
\hspace{5.cm}
\nonumber
\\
\nonumber
 F^{\alpha lm_l}_{\alpha'l'm_l'}(R \rightarrow \infty) 
\sim   \delta_{\alpha \alpha'} \delta_{ll'} 
\delta_{m_lm_l'} 
\exp{[-{\rm i}( k_{\alpha} R - \pi l/2)]}
\\
-  
\left (
\frac{k_{\alpha}}{k_{\alpha'}}
\right )^{1/2}
S_{\alpha'l'm_l'; \alpha lm_l}
\exp{[{\rm i}( k_{\alpha'} R - \pi l'/2)]} 
\label{boundary}
\end{eqnarray}

\noindent
to the solution of the close coupled equations
at a fixed magnetic field strength and total energy $E$
\begin{eqnarray}
\left[\frac{d^2}{dR^2} - \frac{l(l+1)}{R^2} + 2 {\mu} E \right]F_{\alpha lm_l}(R) 
= 2{\mu}\sum_{\alpha'l'm_l'}
 [{\bf C^{\rm T} U C}]_{\alpha lm_l; \alpha'l'm_l'}   
F_{\alpha'l'm_l'}(R).
\label{CC}
\end{eqnarray}

\noindent
The numerical solution of Eqs. (\ref{CC}) with the boundary conditions (\ref{boundary})
yields the scattering $S$-matrix or the probability amplitudes for transitions 
between the collision channels ($\alpha, l, m_l$). The notation $k_\alpha$ is used for 
the wave-number corresponding to channel $\alpha$. 
 The cross sections for elastic and inelastic collisions
are computed from the $S$-matrix as

\begin{eqnarray}
\sigma_{\alpha \rightarrow \alpha'} = \frac{\pi}{k_{\alpha}^2}
 \sum_{l} \sum_{m_l} \sum_{l'} \sum_{m_l'} 
|\delta_{ll'} \delta_{m_lm_l'}  \delta_{\alpha \alpha'}  - S_{\alpha lm_l;\alpha'l'm_l'}|^2.
\label{smat}
\end{eqnarray}

\noindent
The projection of the total angular momentum of the colliding molecules on the magnetic field axis ($M$)
remains a good quantum number in the presence of external fields. That is why the coupling matrix 
${\bf U}$ does not contain couplings between states with different values of $M$
and the transformation ${\bf C}$ does not mix different $M$-states. The equations (\ref{CC})
should therefore be integrated in a cycle over all possible $M$-values and the collision cross section determined 
from the summation over $M$

\begin{eqnarray}
\sigma_{\alpha \rightarrow \alpha'} = \frac{\pi}{k_{\alpha}^2}
\sum_{M}
 \sum_{l} \sum_{m_l} \sum_{l'} \sum_{m_l'} 
|\delta_{ll'} \delta_{m_lm_l'}  \delta_{\alpha \alpha'}  - S^M_{\alpha lm_l;\alpha'l'm_l'}|^2.
\label{smatM}
\end{eqnarray}

In some specific cases (for example, in the presence of strong spin-orbit interactions) it may be 
advantageous to couple internal angular momenta of the molecules with molecular 
spins or it may be easier to determine the matrix elements of the Hamiltonian in the molecule-fixed 
coordinate frame.

\section{External field control of molecular dynamics}

The collision energy of molecules at subKelvin temperatures is less than perturbations 
due to interactions of the molecules with external electric and magnetic fields 
available in the laboratory. External fields may therefore be used to manipulate dynamics (chemical reactions, inelastic collisions and dissociation) of cold and ultracold molecules.  

External field control of molecular dynamics can be based on several different principles. 
Zeeman and Stark effects may remove some of the energetically allowed reaction 
paths or they may open closed reaction channels, leading to suppression or 
enhancement of the reaction efficiency \cite{krems2004}. External fields 
break the spherical symmetry of the space and couple 
the states of different total angular momenta, so that forbidden electronic 
transitions may become allowed in an external field. The transition rate 
may then be controlled by the field strength \cite{krems2003}.  
Some chemical reactions or inelastic collisions are significantly affected by 
centrifugal barriers in the final reaction channels. External fields may
increase the energy separation between the initial and final states
of the reactants and products and enhance the reaction probability to 
a great extent \cite{volpi2002,krems2003,krems2003c}.
External fields may modify intermolecular interaction potentials and induce 
long-range potential minima due to avoided crossings between different electronic states 
of colliding molecules \cite{avdeenkov2003,avdeenkov2004}. 
As a result, molecules may form long range dimer complexes. The binding energy of 
the complexes can be controlled by the strength of the external fields. 
The rate of low temperature abstraction reactions may be dramatically enhanced by 
the presence of a resonance state near threshold \cite{bodo2004}. 
By properly choosing the magnitude of the external field, it should be possible 
to bring an excited bound level of the reaction complex in resonance 
with the collision energy. This will enhance the reaction probability.

  Krems showed that dissociation of weakly bound molecules can be induced by external magnetic fields, if one or more products of the dissociation are in states with non-zero electronic orbital angular momenta \cite{krems2004}. Figure 2 explains the idea of the magnetic-field induced dissociation. 
Several electronic states correlating with different Zeeman levels of the dissociation products have the same asymptotic energy at zero magnetic field. Magnetic fields split the Zeeman energy levels and  bound levels of some electronic states may become embedded in the continuum of the other states in the presence of the field. The molecules can then dissociate through transitions between the Zeeman levels. The Zeeman predissociation may be efficient
and it can be controlled by the magnitude of the external field. As the magnetic field increases, the number of the predissociating  bound levels increases and the predissociation time decreases. Figure 3 shows the lifetime of the He-O($^3P$) van der Waals molecule as a function of the magnetic field.

Volpi and Bohn showed that the probability of  Zeeman relaxation transitions in ultracold molecular collisions is extremely sensitive to the magnitude of the external magnetic field, especially at low fields \cite{volpi2002}. The orbital angular momentum of the colliding molecules about each other is zero in ultracold collisions so there is no centrifugal barrier in the incoming collision channel. 
Due to conservation of the total angular momentum projection, Zeeman transitions must be accompanied by changes of the orbital angular momentum.
 Non-zero orbital angular momenta in final collision channels lead to long-range centrifugal barriers that suppress the inelastic collisions at low 
fields (see Fig. 4). 
In the limit of zero field, the Zeeman levels are degenerate and the cross sections vary with velocity $v$ near threshold as $\sim v^{\Delta m}$ for even 
$\Delta m$ or 
$\sim v^{\Delta m + 1}$ for odd $\Delta m$ \cite{krems2003c}, where $\Delta m$ is the change in the orbital angular momentum projection in the laboratory fixed coordinate system. 
In the presence of the field, the Zeeman relaxation cross sections must vary as $\sim 1/v$ 
according to the Wigner law.  
The energy separation between the Zeeman levels increases with the magnitude of the field so that the centrifugal barriers cannot impede the relaxation dynamics at high fields
 (see Fig. 4). 
Figure 5 presents the zero temperature rate constant for Zeeman relaxation in collisions of rotationally ground-state NH($^3\Sigma$)  molecules in the maximum energy spin level with $^3$He atoms
as a function of the magnetic field.   

  External electric and magnetic fields break the spherical symmetry of the collision problem
  and may couple electronic states otherwise uncoupled \cite{krems2003}. 
  Forbidden electronic transitions may thus become allowed in the presence of external fields. 
  Figure 6 shows the energy diagram of the carbon atom in a magnetic field. 
Krems \etal~\cite{krems2002c} demonstrated that the $^3P_1 \rightarrow ^3P_0$ transition in collisions of carbon with He atoms cannot occur at zero temperature due to the symmetry of the electronic interaction. 
The cross section for this transition vanishes in the limit of zero collision velocity. External magnetic fields couple the $^3P_1$ and $^3P_0$ states of carbon so the cross section for the fine structure relaxation varies as $\sim 1/v$ in the presence of a magnetic field. 
Figure 7 shows the zero temperature rate constant for this forbidden electronic transition as a function of the external  magnetic field. 

  Avdeenkov \etal~\cite{avdeenkov2003,avdeenkov2004} showed that polar molecules may form long-range complexes in the presence of dc electric fields. The complexes are bound in long-range potential wells arising due to avoided crossings of electronic states with repulsive and attractive interactions in the long range.    Electric fields shift the asymptotic energy of the colliding molecules and move the position of the avoided crossings. Thus, electric fields can be used to control the binding energy of the long range complexes and collision properties of ultracold polar molecules. The authors argued that the long-range field-linked states may prevent molecules from approaching to short intermolecular distances where different electronic states are strongly coupled and inelastic relaxation and chemical reactions leading to trap loss are inevitably efficient.   Inelastic collisions and chemical reactions can then be activated by switching off the field.

Bodo \etal~\cite{bodo2004} computed the rate for the F + H$_2$ $\rightarrow$ HF + H chemical reaction as a function of the mass of the hydrogen atoms. Increasing the mass of H$_2$ pushes the least bound level in the entrance reaction channel to above the dissociation threshold and the calculations mimic the chemical reaction near a Feshbach resonance. Figure 8 shows that the zero temperature  reaction rate increases to a great extent near the resonance.  If the reactants and products of a chemical reaction are separated by an activation energy, the chemical reaction occurs at low temperature by tunneling under the potential barrier. Most abstraction chemical reactions have activation barriers and the F + H$_2$ $\rightarrow$ HF + H reaction is a typical example of an abstraction reaction with an activation barrier. 
If the reactants are trapped in a resonance state near the potential barrier, the collision complex lives longer and the tunneling probability is enhanced. 
Feshbach resonances can be induced by external electric or magnetic fields.  So the calculation of Bodo \etal~suggests that abstraction chemical reactions can be induced by tuning Feshbach resonances with external fields. 

 Ticknor and Bohn \cite{ticknor2005} considered the effects of electric and magnetic fields on 
collisions of OH($^2\Pi_{3/2}$) molecules. 
They found that inelastic relaxation of electric-low-field seeking states can be suppressed by magnetic fields. 
The authors showed that an applied magnetic field may counteract some couplings responsible 
for the Stark relaxation and separate the energies of the initial and final scattering channels.  
Magnetic fields of a few thousand Gauss were found to be sufficient to suppress the inelastic relaxation by two orders of magnitude.
Avdeenkov and Bohn \cite{avdeenkov2002} had earlier shown that collisionally induced inelastic relaxation of 
electrostatically trapped polar molecules is very efficient, especially at high electric fields. 
The results of Ticknor and Bohn are, therefore, particularly important as they suggest that 
the trap loss collisions may be mitigated by properly chosen magnetic fields.

\section{Chemical applications of ultracold molecules}

 The field of cold molecules is expanding very quickly. Starting from the experiment on magnetic trapping of CaH molecules \cite{weinstein1998}, it now encompasses the work of more than 
50 research groups worldwide. The interest to cold and ultracold molecules is shared by atomic, molecular and condensed-matter physicists, quantum optics physicists, physical chemists and spectroscopists. 
Why such an interdisciplinary interest?  
 
High precision spectroscopy measurements possible with ultracold molecules may allow for tests of fundamental symmetries of the nature and help in searches for the time variation of the fundamental constants (see the review of Doyle \etal~\cite{editorial} and references therein). Cold trapped molecules can be used in quantum computation \cite{demille2002}.  Molecular Bose-Einstein condensation may lead to novel phenomena yet to be discovered. 

 Possibilities of chemical research with cold and ultracold molecules are boundless and enticing. 
 Particularly appealing are the prospects to explore Bose-enhanced chemistry \cite{ami}. 
 Selectivity of chemical reactions and branching ratios of photodissociation may be greatly enhanced in  a molecular Bose-Einstein condensate due to collective dynamics of condensed molecules.  Selection rules are more strict and propensities for near-resonant energy transfer  in molecular collisions or chemical reactions are more pronounced at ultracold temperatures \cite{krems2002review}.  Collisional relaxation of ultracold molecules may, therefore, be used for the creation of molecules with inverse rotational distributions and molecular lasers.

 As described in this paper, inelastic collisions, chemical reactions and dissociation of cold and ultracold molecules  can be manipulated with external electric and magnetic fields. External field control of molecular dynamics at subKelvin temperatures is an alternative to coherent control methods developed and used in various areas of chemistry \cite{moshe}. Coherent control of bimolecular processes is complicated by the need to entangle internal molecular states with the center-of-mass motion and it will be much easier to implement coherent control schemes with stationary ultracold molecules. 
 
   External field control of molecular dynamics at ultracold temperatures will allow for studies of fine details of molecular structure. Ultracold collisions probe the long range of intermolecular potentials 
and measurements of ultracold dynamics will provide detailed information about weak intermolecular interaction forces. 

  Experiments with ultracold molecules will test the applicability limits of conventional molecular dynamics theories. For example, the Boltzmann equation may not be valid in the ultracold temperature regime as it does not account for quantum effects in molecular interactions. New theories may need to be developed for the proper description of molecular dynamics at temperatures near absolute zero. 
  
  Despite the constantly growing number of research groups working with ultracold molecules, the field of ultracold molecules remains largely a {\it terra incognita}. Every new discovery generates a manifold of questions.  It is safe to say that the field of ultracold molecules has already exploded but it is yet to blossom.

\section*{Acknowledgments}

Thanks are due to John Bohn for useful comments on the manuscript. 
The work of the author is supported by NSF grants to 
the Center for Ultracold Atoms at Harvard University and the Massachusetts Institute of 
Technology and the Institute for Theoretical Atomic, Molecular and Optical Physics at the Harvard-Smithsonian 
Center for Astrophysics.

\clearpage
\newpage

\begin{figure}[ht]
\caption{
Schematic diagram showing variation of the potential energy of atoms or molecules in low-field seeking
states (full line) and high-field seeking states (broken line) in a magnetic trap. 
The minimum of the field strength corresponds to the center of the trap. 
}
\vspace{0.5cm}
\label{fig:1}
\begin{center}
\includegraphics[scale=0.4]{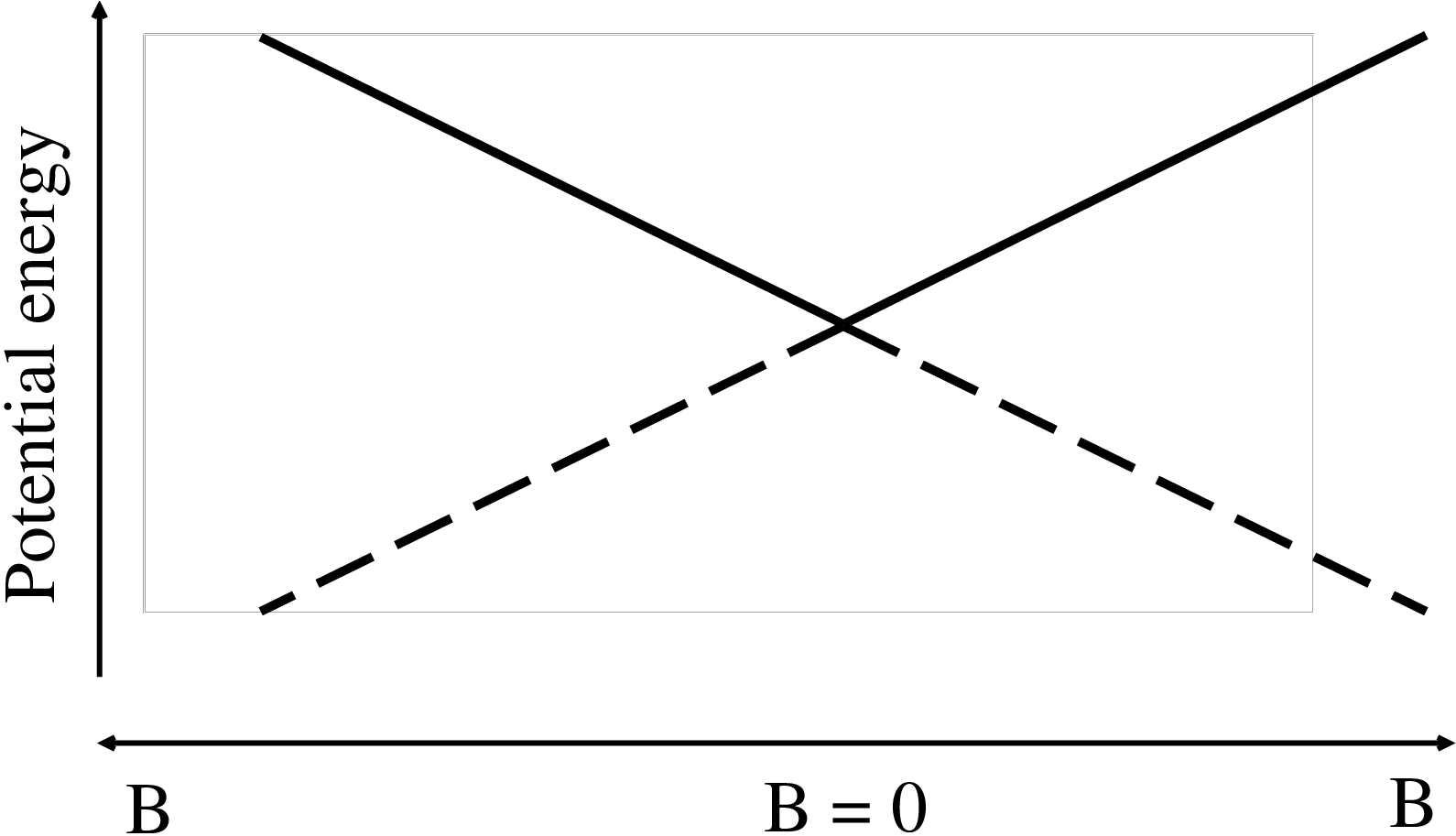}
\end{center}
\end{figure}

\begin{figure}[ht]
\caption{
Magnetic-field induced dissociation. Two electronic states shown by full and broken curves correspond to 
different Zeeman levels of the dissociation products and they are asymptotically degenerate at zero field. 
The asymptotic energies of the electronic states separate in the presence of the field and bound energy 
levels of one state may become embedded in the continuum of the other state. The molecules may then dissociate 
through Zeeman transition. Adapted from Ref. \cite{krems2004}. 
}
\label{fig:2}
\begin{center}
\includegraphics[scale=0.7]{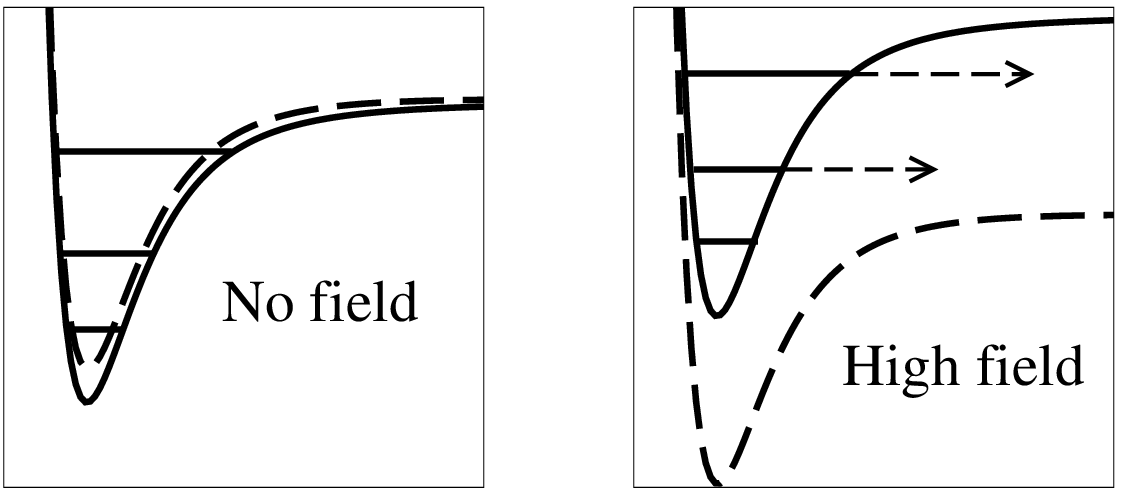}
\end{center}
\end{figure}

\begin{figure}[ht]
\caption{
Lifetime of the He-O($^3P$) van der Waals complex as a function of the external 
magnetic field. Adapted from the data of Ref. \cite{krems2004}.
}
\label{fig:3}
\begin{center}
\includegraphics[scale=0.37]{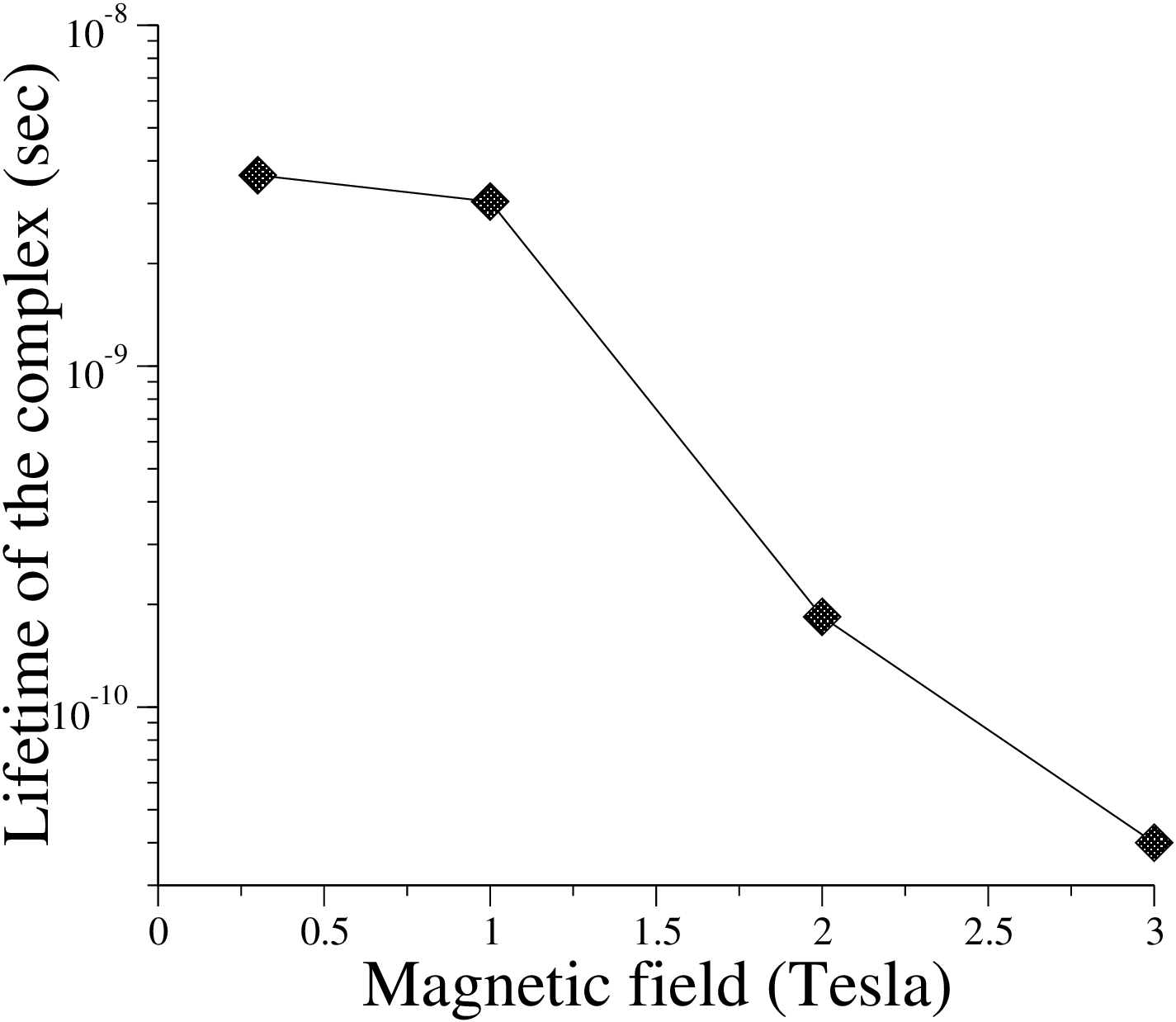}
\end{center}
\end{figure}

\begin{figure}[ht]
\caption{
External field suppression of the role of centrifugal barriers in outgoing reaction channels. 
Incoming channels are shown by full curves, outgoing channels --  by broken curves. 
An applied field separates the energies of the initial and final channels and suppresses the 
role of the centrifugal barriers in the outgoing channels. 
}
\label{fig:4}
\begin{center}
\includegraphics[scale=0.57]{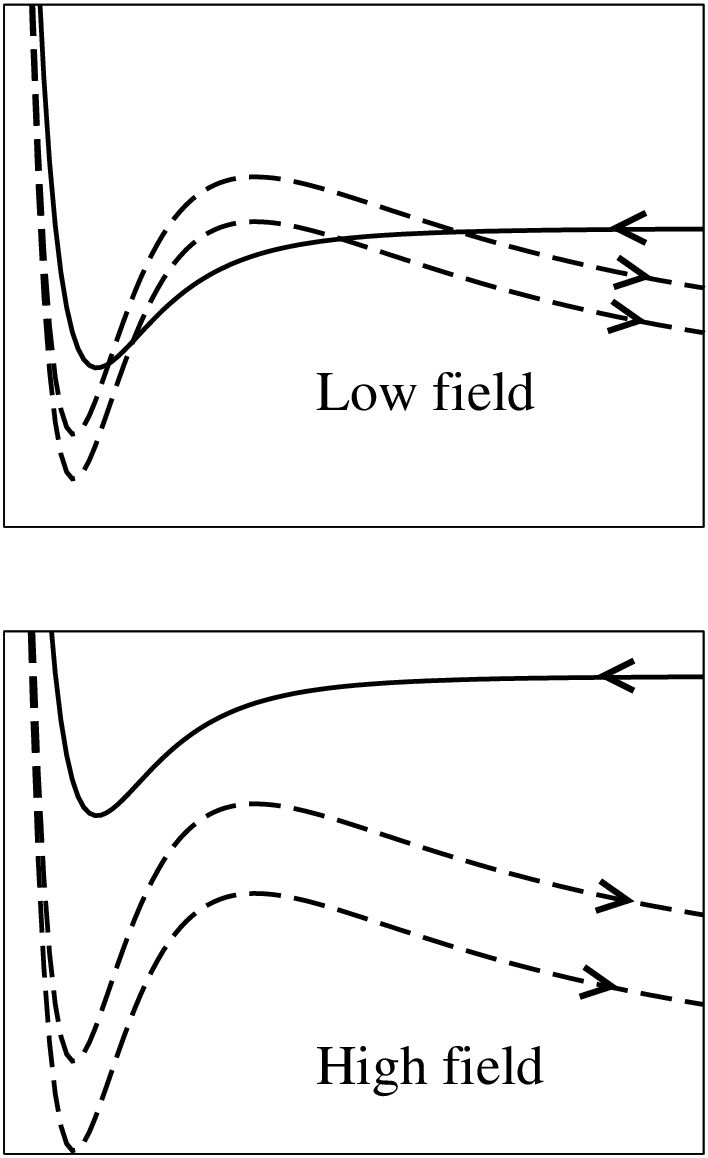}
\end{center}
\end{figure}

\begin{figure}[ht]
\caption{
Zero temperature rate constant for Zeeman relaxation in collisions of rotationally 
ground-state NH($^3\Sigma$) molecules in the maximally stretched spin level  with $^3$He atoms.
Such field dependence is typical for Zeeman or Stark relaxation in ultracold collisions of atoms 
and molecules without hyperfine interaction. The variation of the relaxation rates with 
the field is stronger and extends to larger field values for systems with smaller reduced mass. 
See Fig. 4. 
}
\vspace{1.cm}
\label{fig:5}
\begin{center}
\includegraphics[scale=0.37]{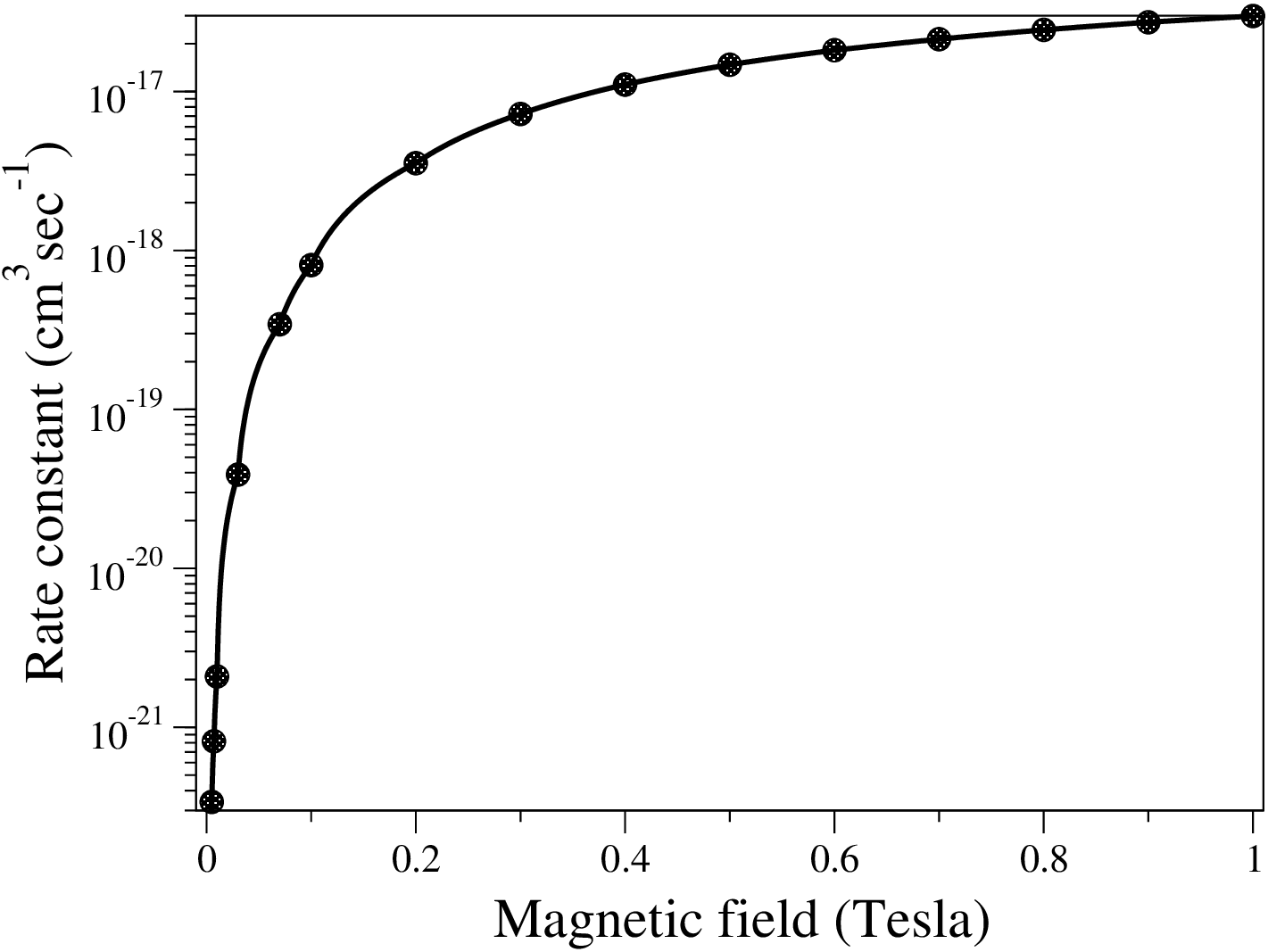}
\end{center}
\end{figure}

\begin{figure}[ht]
\caption{
Energy levels of the carbon atom in a magnetic field. 
}
\label{fig:6}
\begin{center}
\includegraphics[scale=0.47]{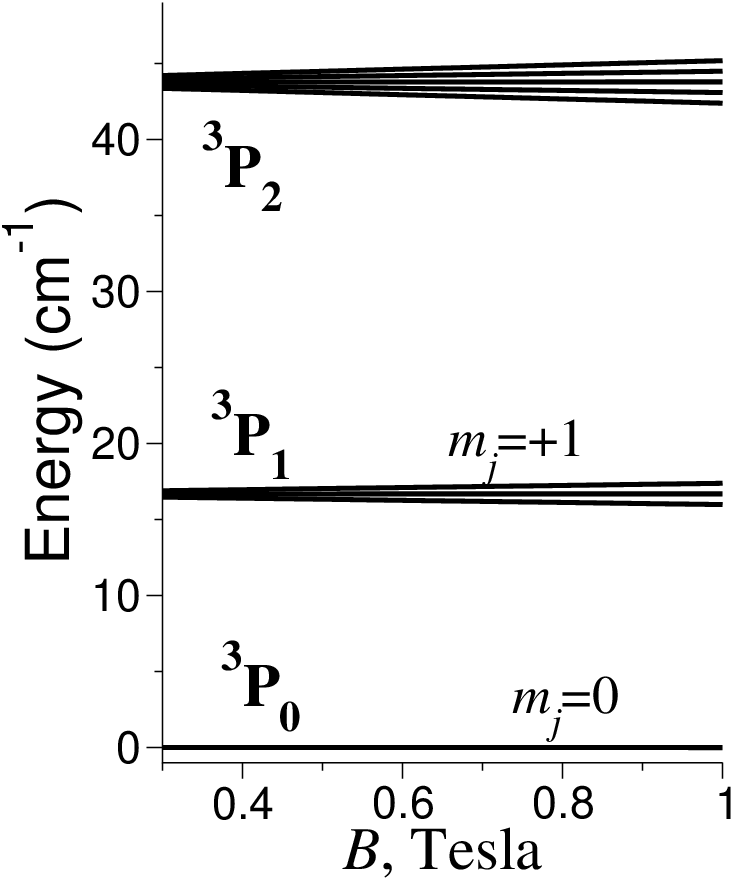}
\end{center}
\end{figure}

\begin{figure}[ht]
\caption{
Zero temperature rate constant for the forbidden $^3P_1(m_j=+1) \rightarrow ^3P_0$ transition in collisions of
carbon atoms with $^3$He. The magnetic field couples the $^3P_1$ and $^3P_0$ states and induces 
the inelastic transition. Adapted from the data of Ref. \cite{krems2003}. 
}
\label{fig:7}
\begin{center}
\includegraphics[scale=0.37]{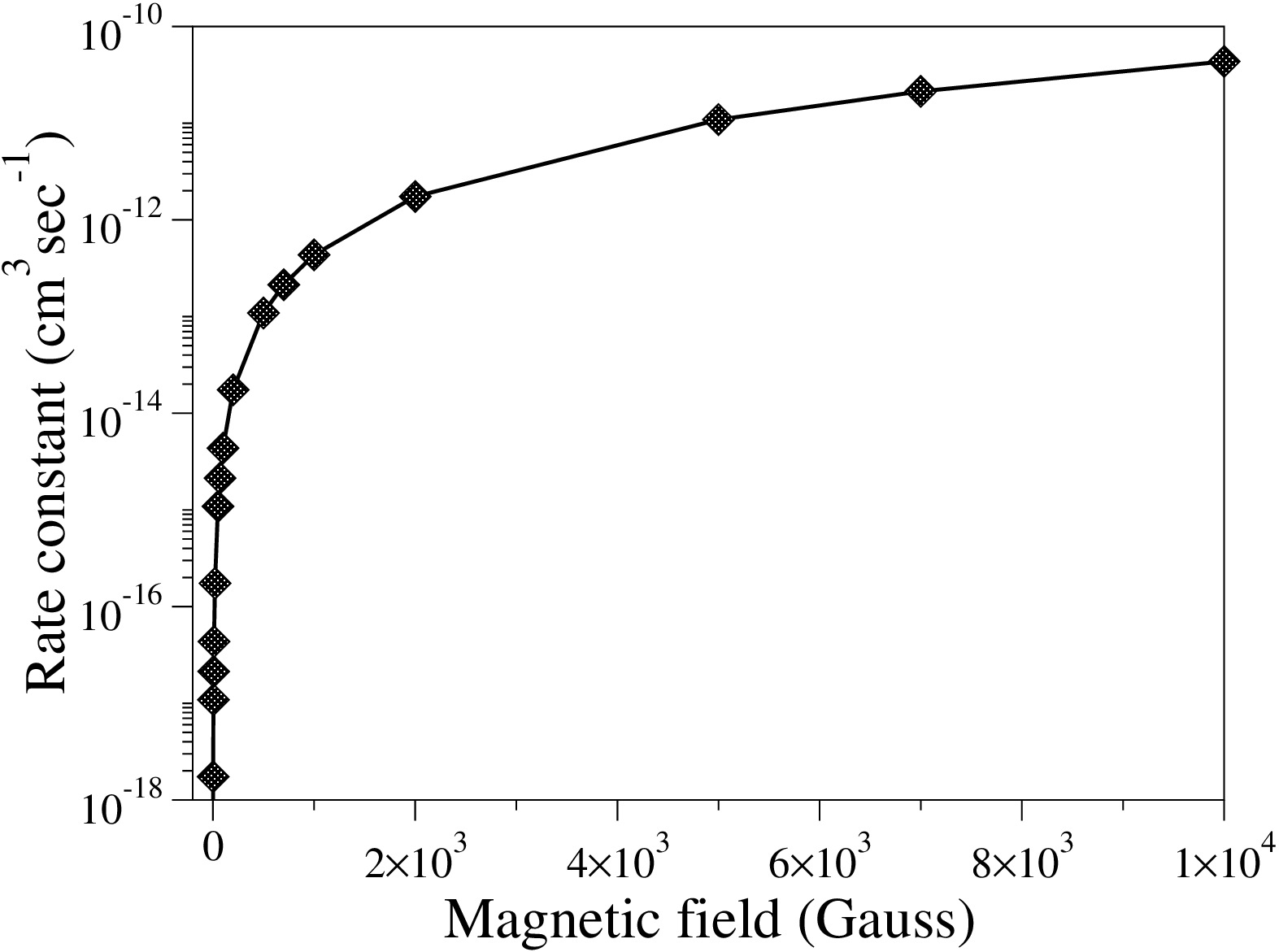}
\end{center}
\end{figure}

\begin{figure}[ht]
\caption{
Zero temperature rate constant for the 
F + H$_2$ $\rightarrow$ HF + H chemical reaction. 
The presence of the resonance near threshold enhances the reaction rate to a great extent. 
Adapted from Ref. \cite{bodo2004}. 
}
\label{fig:8}
\begin{center}
\includegraphics[scale=0.37]{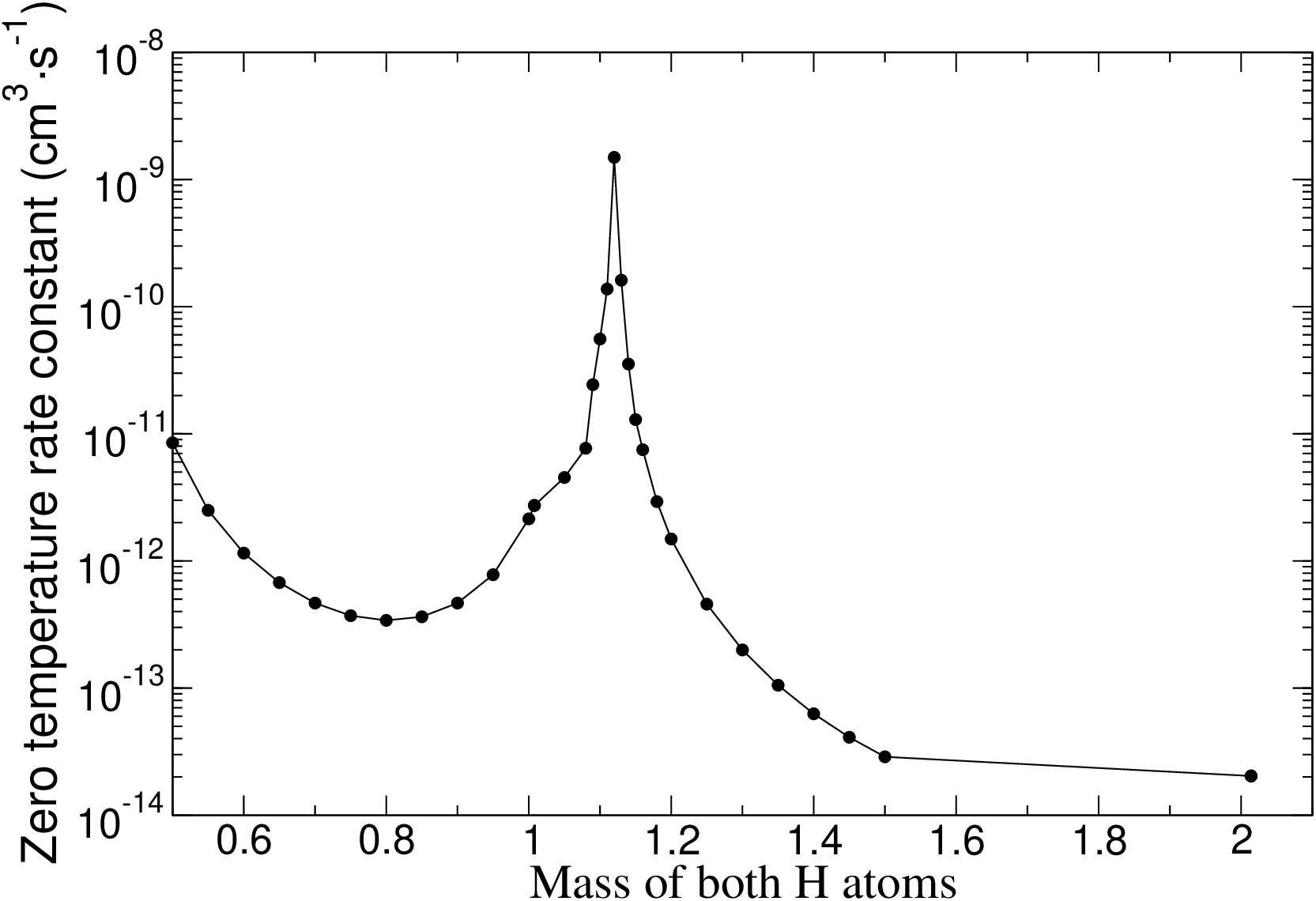}
\end{center}
\end{figure}

\clearpage
\newpage

\begin{table}
\caption{
The coldest molecules produced in laboratory studies to date. 
$T$ is the lowest temperature and $N$ is the maximum 
number of the molecules 
achieved in the indicated type of the experiments. 
}
\label{table1}
\begin{ruledtabular}
\begin{tabular}{c c c c c}

Method   & Molecule  &  $T$  &  $N$  \\
\hline

Feshbach resonance & Li$_2$, K$_2$, Cs$_2$, Rb$_2$, Cs$_4$         & $50$ nK            & 900,000    \\

Photoassociation        &  Rb$_2$, Cs$_2$, He$^*_2$, H$_2$,  &      $25$ $\mu$K  &  200,000   \\
                                       &  Li$_2$, Na$_2$, K$_2$, Cs$_2$,        &     \\
                                       &    KRb, RbCs                                              &                               &                  \\
Stark Deceleration     &   $^{14}$NH$_3$, $^{14}$ND$_3$, $^{15}$NH$_3$,    &    $25$ mK       &   10,000  \\ 
                                       &      CO, OH, YbF                                         &    \\
Skimming                     &         H$_2$CO, ND$_3$, S$_2$            &      1 K                 &    --               \\
Crossed beam            &                         NO                                        &      $400$ mK   &    --               \\
Buffer-gas loading     &        CaH, CaF, VO, PbO, NH,                   &       $400$ mK  &       $10^{13}$  \\

\end{tabular}
\end{ruledtabular}
\end{table}

\begin{table}
\caption{
Ratio $\gamma$ of rate constants for elastic and inelastic (trap loss) collisions
at $T=1.8$ K and $B = 3.8$ T.
}
\label{table2}
\begin{ruledtabular}
\begin{tabular}{c c c c}
\hline

Collision complex   &  $\gamma$  \footnote{The theoretical value from Ref. \cite{krems2005}}
 & $\gamma_{\rm exp}$ \footnote{The experimental value from Ref. \cite{cindy}}\\
\hline

 O($^3P$)-$^3$He    &  3     & \\
 O($^1D$)-$^3$He    &  1.6   & \\
 Sc($^2D$) - $^3$He & 790    & $< (1.6 \pm 0.3) \times 10^{4}$ \\
 Ti($^3F$) - $^3$He & 6953   &  $\sim (4.0 \pm 1.8) \times 10^{4}$ \\
\end{tabular}
\end{ruledtabular}
\end{table}


\clearpage
\newpage






\clearpage
\newpage


\begin{thebibliography}{99}


\bibitem{note1}
High temperature chemical reactions of radicals in solutions can be controlled by external static and oscillating magnetic fields. For a particular example, see the work of R. J. Woodward, C. R. Timmel, P.J. Hore, and K. McLaughlan, Mol. Phys. {\bf 100}, 1181 (2002). This type of external field control is based on inducing interconversion between singlet and triplet states of the reactive complexes and it should be possibile in gas phase reaction dynamics as well.

\bibitem{editorial}
J. Doyle, B. Friedrich, R.V. Krems, and F. Masnou-Seeuws,
Eur. Phys. J. D {\bf  31}, 149 (2004). 


\bibitem{note2}
For collisions of identical fermionic molecules it is one.


\bibitem{wigner}
E.P.  Wigner,  Phys. Rev. {\bf 73}, 1002 (1948).

\bibitem{krems2004}
R.V. Krems,  \prl{93}{013201}{2004}.

\bibitem{krems2003}
R.V. Krems  and A. Dalgarno,  
\pra{68}{013406}{2003}.

\bibitem{bodo2004}
E. Bodo, F.A. Gianturco,  N. Balakrishnan,  and A. Dalgarno,
\jpb{37}{3641}{2004}.


\bibitem{krems2002review}
R.V. Krems,  in  ``{\it Recent Research Developments in Chemical Physics}",
Vol. 3, 485 (2002).

\bibitem{bethlem2003}
H.L. Bethlem  and  G. Meijer, 
Int. Rev. Phys. Chem. {\bf 22}, 73 (2003).


\bibitem{hossein}
H.R. Sadeghpour, J.L. Bohn, M.J. Cavagnero, B.D. Esry, 
I.I. Fabrikant, J.H. Macek, and A.R.P. Rau,
\jpb{33}{R93}{2000}.

\bibitem{landau}
L.D. Landau,  and E.M. Lifshitz,  {\it Quantum mechanics}, (Nauka, Moscow, 1965).


\bibitem{mott}
N.F. Mott and H.S.W. Massey,  ``{\it The Theory of Atomic Collisions}'', 
3rd ed. (Clarendon, Oxford, 1965), p. 380.

\bibitem{bala1997}
N. Balakrishnan, V. Kharchenko, R.C. Forrey, and A. Dalgarno,   
\cpl{280}{5}{1997}.

\bibitem{bohn1997}
J.L. Bohn and P.S. Julienne, 
\pra{56}{1486}{1997}.


\bibitem{bala1998}
N. Balakrishnan, R.C. Forrey, and A. Dalgarno,  
\prl{80}{3224}{1998}.


\bibitem{forrey1999}
R.C. Forrey, N. Balakrishnan,  A. Dalgarno,
 M.R. Haggerty, and  E.J. Heller, 
\prl{82}{2657}{1999}.


\bibitem{bala2000}
N. Balakrishnan, R.C. Forrey, and A. Dalgarno,  
\jcp{113}{621}{2000}.

\bibitem{bala2001}
N. Balakrishnan, and A. Dalgarno,  
\jpc{105}{2348}{2001}.


\bibitem{zhu2001}
C. Zhu, N.  Balakrishnan, and A. Dalgarno, 
\jcp{115}{1335}{2001}.


\bibitem{flasher2002}
J.C. Flasher and R.C. Forrey, 
\pra{65}{032710}{2002}.


\bibitem{bodo2002}
E. Bodo, F.A. Gianturco, and A. Dalgarno,  
\cpl{353}{127}{2002}.


\bibitem{bala2003}
N. Balakrishnan,  G.C. Groenenboom, R.V.  Krems,  and A. Dalgarno,   
\jcp{118}{7386}{2003}.


\bibitem{florian2004}
P. Florian, M. Hoster,  and R.C. Forrey,  
\pra{70}{032709}{2004}.

\bibitem{tilford2004}
K. Tilford, M. Hoster, P.M. Florian, and R.C. Forrey,   
\pra{69}{052705}{2004}.


\bibitem{krems2001}
R.V. Krems and A.A. Buchachenko,   
\pra{64}{024704}{2001}.


\bibitem{siska2001}
P.E. Siska, 
\jcp{115}{4527}{2001}.


\bibitem{krems2002}
R. Krems and A. Dalgarno, 
\jcp{117}{118}{2002}.

\bibitem{krems2002a}
R.V. Krems and A. Dalgarno,   
\pra{66}{012702}{2002}.



\bibitem{bala2001a}
N. Balakrishnan and A. Dalgarno,  
\cpl{341}{652}{2001}.

\bibitem{bala2003a} 
N. Balakrishnan and A. Dalgarno,   
\jpc{107}{7101}{2003}.


\bibitem{bodo2002a}
E. Bodo, F.A. Gianturco, and A. Dalgarno,   
\jpb{35}{2391}{2002}.

\bibitem{bodo2002b}
E. Bodo, F.A. Gianturco, and A. Dalgarno,   
\jcp{116}{9222}{2002}.

\bibitem{soldan2002}
P. Sold\'{a}n, M.T.  Cvita\v{s}, J.M. Hutson,
P. Honvault, and  J.-M. Launay,  
\prl{89}{153201}{2002}.


\bibitem{cvitas2005}
M.T. Cvita\v{s}, P.  Sold\'{a}n, J.M.  Hutson, P. Honvault, and J.-M. Launay, 
\prl{94}{033201}{2005}.

\bibitem{cvitas2005a}
M.T. Cvita\v{s}, P.  Sold\'{a}n, J.M.  Hutson, P. Honvault, and J.-M. Launay, 
Phys. Rev. Lett, in press; 
ArXiv: cond-mat/0501636. 

\bibitem{quemener2005}
G. Qu\'{e}m\'{e}ner, P. Honvault, J.-M. Launay,   P. Sold\'{a}n, D.E. Potter, and  J.M. Hutson, 
Phys. Rev. A, in press; 
ArXiv:cond-mat/0411158.


\bibitem{ketterle}
W. Ketterle,
 \rmp{74}{1131}{2002}.


\bibitem{phillips}
W.D. Phillips, \rmp{70}{721}{1998}.


\bibitem{doyle1995}
J.M. Doyle, B. Friedrich, J. Kim,  and D. Patterson,
 \pra{52}{R2515}{1995}.

\bibitem{weinstein2002}
J.D. Weinstein, R. deCarvalho, C.I. Hancox,  and J.M. Doyle,
 \pra{65}{021604}{2002}.

\bibitem{dirosa2004}
 M.D. Di Rosa, Eur. Phys. J. D {\bf 31}, 395 (2004). 

\bibitem{weinstein1998}
 J. Weinstein,  R. deCarvalho, T. Guillet, B. Friedrich,
and  J.M. Doyle,   
Nature (London), {\bf 395}, 148 (1998).


 \bibitem{rangwala2003}
S.A. Rangwala, T. Junglen, T. Rieger, P.W.H. Pinkse, and G. Rempe, 
\pra{67}{043406}{2003}.


\bibitem{nikitin2003}
E. Nikitin,  E. Dashevskaya, J. Alnis,   
M. Auzinsh,  E.R.I. Abraham,  B.R. Furneaux, M. Keil, C. McRaven, N. Shafer-Ray,  and R.  Waskowsky,
\pra{68}{023403}{2003}.


\bibitem{gupta1999}
M. Gupta and D. Herschbach,  
J. Phys. Chem. A {\bf 103}, 10670 (1999).


\bibitem{elioff2003}
M.S. Elioff,  J.J. Valentini, and D.W. Chandler, 
Science {\bf 302}, 1940 (2003).

\bibitem{bohn2000}
J.L. Bohn, 
\pra{61}{040702(R)}{2000}.

\bibitem{bohn2000a}
J.L. Bohn,  
\pra{62}{032701}{2000}.

\bibitem{bohn2001}
J.L. Bohn, 
\pra{63}{052714}{2001}.

\bibitem{avdeenkov2001}
A.V.  Avdeenkov,  and J.L. Bohn,   
\pra{64}{052703}{2001}.

\bibitem{volpi2002}
A. Volpi and J.L. Bohn,  
\pra{65}{052712}{2002}.

\bibitem{volpi2002a}
A. Volpi and J.L. Bohn,  
\pra{65}{064702}{2002}.


\bibitem{krems2003a}
R.V. Krems, A. Dalgarno, N. Balakrishnan, and 
G.C. Groenenboom,
\pra{67}{060703(R)}{2003}.


\bibitem{krems2003b}
R.V. Krems, H.R. Sadeghpour, A. Dalgarno, D. Zgid, J. Klos, and G. Chalasinski, 
\pra{68}{051401(R)}{2003}.


\bibitem{krems2004a}
R.V. Krems and A. Dalgarno, 
\jcp{120}{2296}{2004}.


\bibitem{groenenboom2005}
G.C. Groenenboom,  private communication. 


\bibitem{maussang2005}
K. Maussang, D. Egorov, J.S.  Helton, S.V. Nguyen,  and  J.M. Doyle,   
Phys. Rev. Lett., in press.

\bibitem{meijer}
G. Meijer, private communication. 

\bibitem{soldan2004}
P. Sold\'{a}n and  J.M. Hutson,  
\prl{92}{163202}{2004}.


\bibitem{greiner2003}
M. Greiner, C.A. Regal, and D.S. Jin,   
Nature (London) {\bf 426}, 537 (2003).


\bibitem{jochim2003}
S. Jochim, M. Bartenstein, A. Altmeyer, G. Hendl, S. Riedl, C. Chin,  J.H. Denschlag,   
and R. Grimm, 
Science {\bf 302}, 2101 (2003).


\bibitem{zwierlein2003}
M.W. Zwierlein,  C.A. Stan, C.H.  Schunck, S.M.F.  Raupach,   
S. Gupta, Z. Hadzibabic, and  W. Ketterle,  
\prl{91}{250400}{2003}.


\bibitem{petrov2003}
D.S. Petrov,  
 \pra{67}{010703(R)}{2003}.

\bibitem{petrov2004}
D.S. Petrov, C. Salomon,  and  G.V. Shlyapnikov,   
\prl{93}{090404}{2004}.
 

\bibitem{soldan2003}
P. Sold\'{a}n,  M.T. Cvita\v{s}, and  J.M. Hutson,   
\pra{67}{054702}{2003}.


\bibitem{bergeman2004}
T. Bergeman, A.J. Kerman, J. Sage, S. Sainis, and D. DeMille, 
Eur. Phys. J. D {\bf 31}, 179 (2004). 

\bibitem{wang2004}

D. Wang, J. Qi, M.F. Stone, O. Nikolayeva, B. Hattaway, S.D. Gensemer, H. Wang. W.T. Zemke, P.L. Gould,  E.E. Eyler, and W.C. Stwalley,  
 Eur. Phys. J. D {\bf 31}, 165 (2004). 


\bibitem{ticknor2005}
C. Ticknor and J.L. Bohn, \pra{71}{022709}{2005}. 


\bibitem{avdeenkov2002}
A.V. Avdeenkov  and J.L. Bohn,   
\pra{66}{052718}{2002}.


\bibitem{zygelman1994}
B. Zygelman, A. Dalgarno, and  R.D. Sharma,  
\pra{49}{2587}{1994}.


\bibitem{krems2002b}
R.V. Krems, A.A. Buchachenko, M.M. Szczesniak, J. Klos,  and G. Chalasinski,   
\jcp{116}{1457}{2002}.


\bibitem{krems2004b}
R.V. Krems, G.C. Groenenboom, and A. Dalgarno,   
J. Phys. Chem. A {\bf 108}, 8941 (2004). 


\bibitem{aquilanti}
V. Aquilanti and G. Grossi,  
\jcp{73}{1165}{1980}. 



\bibitem{kokoouline2003}
V. Kokoouline, R. Santra, and C.H. Greene,  
\prl{90}{253201}{2003}.


\bibitem{cindy}
C.I. Hancox, S.C. Doret, M.T. Hummon, R.V. Krems, and J.M. Doyle,
\prl{94}{013201}{2005}.

\bibitem{krems2005}
R.V.  Krems, J. Klos, M.F. Rode, M.M.  Szczesniak, G. Chalasinski,  and A. Dalgarno,   
\prl{94}{013202}{2005}.

\bibitem{jacek}
J. Klos, M.F. Rode, J.E. Rode, G. Chalasinski, and M.M. Szczesniak, 
Eur. Phys. J. D {\bf 31}, 439 (2004). 

\bibitem{cindy_nature}
C.I. Hancox, S.C. Doret, M.T. Hummon, L. Luo, and J.M. Doyle,   
Nature (London) {\bf 431}, 281 (2004). 


\bibitem{krems2004c}
R.V. Krems and A. Dalgarno,  in ``{\it Fundamental world of quantum chemistry}", eds: E. J. Br\"{a}ndas and E. S. Kryachko, vol. III, p. 273 (Kluwer, 2004).


\bibitem{alex}
 A.M. Arthurs  and  A. Dalgarno,  Proc. R. Soc. London, 
Ser. A,  {\bf 256}, 540 (1960). 
 

\bibitem{krems2003c}
R.V. Krems and A. Dalgarno,  
\pra{67}{050704}{2003}.


\bibitem{avdeenkov2003}
A.V. Avdeenkov  and J.L. Bohn,   
\prl{90}{043006}{2003}.


\bibitem{avdeenkov2004}
A.V. Avdeenkov, D.C.E. Bortolotti, and  J.L. Bohn,  
 \pra{69}{012710}{2004}. 


\bibitem{krems2002c}
R.V. Krems, D. Zgid, G. Chalasinski,  J. Klos, and A. Dalgarno,   
\pra{66}{030702(R)}{2002}.


\bibitem{demille2002}
D. DeMille,  \prl{88}{067901}{2002}.


\bibitem{ami}
M.G. Moore and A. Vardi,  \prl{88}{160402}{2002}. 


\bibitem{moshe}
M. Shapiro and P. Brumer,   ``{\it Principles of Quantum Control of Molecular Processes}" (John Wiley and Sons, Inc., New Jersey, 2003). 




\end{thebibliography}
\end{document}